\documentclass[12pt]{article}
\usepackage{graphicx}
\usepackage{caption}
\usepackage{amssymb}
\usepackage[numbers,sort&compress,square]{natbib}
\setcitestyle{numbers}
\setcitestyle{square}
\usepackage{url}
\usepackage{color}
\setcitestyle{numbers}
\setcitestyle{square}
\usepackage{url}
\usepackage{color}

\def\be{\begin{eqnarray}}
\def\ee{\end{eqnarray}}
\def\({\left(}
\def\){\right)}

\usepackage[ddmmyyyy]{datetime}
\def\be{\begin{equation}}
\def\ee{\end{equation}}

\def\({\left(}
\def\){\right)}

\begin{document}
\begin{center}
\Large{Comment on 'Phase transition temperatures of 405-725 K in superfluid ultra-dense 
hydrogen clusters on metal surfaces' [AIP Advances 6, 045111 (2016)]}\\
\vspace{1cm}
\Large{Klavs Hansen}\footnote{klavshansen@tju.edu.cn,  hansen@lzu.edu.cn }\\
\normalsize{Lanzhou Center for Theoretical Physics, Key Laboratory of Theoretical Physics 
of Gansu Province, Lanzhou University, Lanzhou, Gansu 730000, China}\\
and\\
\normalsize{Center for Joint Quantum Studies and Department of Physics, 
School of Science, Tianjin University, 92 Weijin Road, Tianjin 300072, China}\\

\vspace{0.5cm}
\Large{Jos Engelen}\\
\normalsize{Faculty of Science, University of Amsterdam\\ 
Nikhef, Science Park 105, 1098 XG Amsterdam, Netherlands}
\end{center}

\vspace{0.3cm}
The article in \cite{HolmlidAIPA2016} has recently come to our attention. 
It makes a number of extraordinary claims that each contradict 
well established facts in several fields, ranging from atomic and molecular physics to 
superconductivity and superfluidity, with practically no supporting evidence.
We think it worthwhile to rectify the literature with this comment. 

Already the title of the paper announces a couple of truly revolutionary discoveries, 
if they had been correct.
Superfluidity has to date been observed in liquid helium-4 below the lambda 
point of 2.17 K at ambient pressure \cite{Schmitt2015}.
Even below that temperature the superfluid fraction is only reaching 50 \%  around 1.9 K.
For the helium-3 isotope superfluidity occurs around 2 mK \cite{Schmitt2015}. 
The paper of Holmlid and Kotzias makes the remarkable claim of a range of transition 
temperatures between 405 K and 725 K.

The claim is thus that the quantum effects observed by the scientific community 
over a period of more than a century, with the Herculean effort starting with the 
production of liquid helium by Kamerlingh Onnes, should exist at temperatures exceeding 
the melting points of lead.

Also a transition temperature to superconductivity is claimed to be bracketed by these
values. 
This is a truly remarkable statement, even if the claim were only that hydrogen is 
superconducting at all, not the least if this is measured in the time-of-flight 
experiments described.

An inspection of the evidence presented by Holmlid and Kotzias does not 
in our opinion warrant any conclusions of such truly historical magnitude.
Or in fact of any value for a superconducting and superfluid transition at all.
The experimental results presented by the authors consist of four time-of-flight 
spectra.
They are all of rather poor resolution, with large amounts of unresolved intensity 
at late times.
These later peaks have some structure but none that can be identified with any 
kind of  certainty, and their widths are on the order of their mean flight times.
Even the single peak that is seen around 500 ns flight time is not particularly well
resolved.
It is simply not possible to conclude much from data of this quality.
In particularly any conclusion that some superfluid state exists on the substrate
is completely detached from the experimental data presented.
In our experience the spectra rather look like the manifestation of a charging effect 
in the equipment, although other instrumental artifacts may also contribute.
We think some care should have been exercised by the authors before adopting 
the boldest possible interpretation for the spectra.

The authors reiterate their interpretation of their poorly resolved spectra 
as representing the kinetic energies released in a Coulomb explosion and their
claim that this energy appears as the result of the breakup of pairs of point charges.
They then use the observed energies as a measure of the distance between these 
two charges (see their Eq. 1). 
Apart from the experimentally very shaky ground on which this claim is based, 
it is close to impossible to reconcile with energy conservation. 
The authors claim an energy release of 640 eV of their postulated compounds,
based on this argument.
This is supposed to be the electrostatic repulsion between two hydrogen nuclei. 
If this matter were stable in the neutral state before exposure to the laser pulse,
as claimed, this positive energy should be compensated by at least the same amount of 
negative energy supplied by the electrons. 
Otherwise the compound would not be stable and in particular definitely not the ground state 
as claimed by the authors.
The authors' assignment requires that the electrons absorb at least 
640 eV in order to produce the charged particles.
The authors fail to explain how an energy of this magnitude is imparted to the 
molecule.
In fact, they even explicitly state that the electrons are easily removed by the 
laser pulse, and seemingly ignore the question entirely.

The extremely well tested and well understood quantum theory of matter, 
and atoms and molecules in particular, excludes any behavior of the kind 
suggested by Holmlid et al. 
For example, the hydrogen molecule is known from spectroscopy to have an 
equilibrium length of 0.74 \AA \cite{HertzbergDiatomic}.
Any reduction of this distance to the proposed 2 picometer grossly contradicts
all this knowledge.
This should even be clear from an understanding of quantum mechanics at 
an elementary level \cite{AtkinsSixth}.

But even disregarding the arguments based on the huge literature on molecular 
structure and quantum mechanical arguments, one can easily see that the claimed 
molecular structures are not the stable ground state phase of hydrogen.
Hydrogen has been liquified for decades and its phase diagram is well known. 
Nowhere does a phase with the claimed extremely high density appear.

We must also point out that the postulated existence of the ultradense hydrogen 
considered by the authors as a form of Rydberg matter is a rather blatant 
contradiction in terms.
The word Rydberg implies that atoms have been excited to high-lying states, with 
their concomitant large principal quantum numbers and sizes. 
Principal quantum numbers of at least four or five are involved.  
The known large physical dimensions of Rydberg states are now 
postulated by the authors, without any argument, to be converted into extremely 
small values, more than an order of magnitude \textit{smaller} than the Bohr 
radius.
Everything known about the behavior of the hydrogen atom and its quantum 
states, as well as its binding to other atoms, disagrees with this postulated behavior.
In addition, the high orbital quantum numbers responsible for Rydberg states are 
willy-nilly eliminated by the authors who instead ascribe these states to spin 
excitations with total spins of a few units of Planck's constant. 

In summary, the cited paper contains a string of incredible claims about new 
physics.
The claims are based on evidence which with the best of wills can only be 
considered very flimsy, and can only be made with a complete disregard of 
well established scientific facts.

KH acknowledges financial support from the National Science Foundation of China 
with grant No. 12047501 and the 111 Project with No. B20063 from the Ministry of 
Science and Technology of People's Republic of China. 

The authors have no conflicts of interest to declare.

\end{document}